\documentclass[acmlarge]{jair} 

\setcopyright{cc}
\copyrightyear{2025}
\acmDOI{10.1613/jair.1.xxxxx}

\JAIRAE{Daniel Harabor, Sven Koenig}
\JAIRTrack{MAPF} 
\acmVolume{4}
\acmArticle{6}
\acmMonth{8}
\acmYear{2025}

\usepackage{amsmath,amsfonts}
\usepackage{amsthm}
\usepackage{stfloats}
\usepackage{float}
\usepackage{url}
\usepackage{array}
\usepackage[caption=false,font=normalsize,labelfont=sf,textfont=sf]{subfig}
\usepackage[ruled,vlined]{algorithm2e}
\usepackage{subcaption}
\usepackage{graphicx}

\usepackage{mathrsfs} 

\newtheorem{theorem}{Theorem }
\newtheorem{lemma}{Lemma}

\newtheorem{definition}{Definition}

\newcommand{\nodeset}{\mathbb{V}}
\newcommand{\edgeset}{\mathbb{E}}

\begin{document}

\title[\textsc{Dynamic-hjsg}]{Efficient Multi-Agent Coordination via Dynamic Joint-State Graph Construction}

\author{Yanlin Zhou}
\authornote{Corresponding Author.}
\orcid{0000-0002-6820-6772}
\email{yzhou30@gmu.edu}
\affiliation{%
  \institution{George Mason University}
  \city{Fairfax}
  \state{Virginia}
  \country{United States of America}
}

\author{Manshi Limbu}
\orcid{0009-0007-7560-7607}
\email{klimbu2@gmu.edu}
\affiliation{%
  \institution{George Mason University}
  \city{Fairfax}
  \state{Virginia}
  \country{United States of America}
}

\author{Xuesu Xiao}
\orcid{0000-0001-5151-2186}
\email{xiao@gmu.edu}
\affiliation{%
  \institution{George Mason University}
  \city{Fairfax}
  \state{Virginia}
  \country{United States of America}
}


\renewcommand{\shortauthors}{Zhou et al.}

\begin{abstract}
{\bf Background:} 
    Multi-agent pathfinding (\textsc{MAPF}) traditionally focuses on collision avoidance, but many real-world applications require active coordination between agents to improve team performance. 
    
    {\bf Objectives:}
    This paper introduces Team Coordination on Graphs with Risky Edges (\textsc{tcgre}), where agents collaborate to reduce traversal costs on high-risk edges via support from teammates. We reformulate \textsc{tcgre} as a 3D matching problem—mapping robot pairs, support pairs, and time steps—and rigorously prove its NP-hardness via reduction from Minimum 3D Matching.
    
    {\bf Methods:}
    To address this complexity, (in the conference version) we proposed efficient decomposition methods, reducing the problem to tractable subproblems:
    Joint-State Graph (\textsc{jsg}): Encodes coordination as a single-agent shortest-path problem.
    Coordination-Exhaustive Search (\textsc{ces}): Optimizes support assignments via exhaustive pairing.
    Receding-Horizon Optimistic Cooperative A* (\textsc{rhoca}*): Balances optimality and scalability via horizon-limited planning.
    Further in this extension, we introduce a dynamic graph construction method (\textsc{Dynamic-hjsg}), leveraging agent homogeneity to prune redundant states and reduce computational overhead by constructing the joint-state graph dynamically. 
    
    {\bf Results:}
    Theoretical analysis shows \textsc{Dynamic-hjsg} preserves optimality while lowering complexity from exponential to polynomial in key cases. Empirical results validate scalability for large teams and graphs, with HJSG outperforming baselines greatly in runtime in different sizes and types of graphs.
    
    {\bf Conclusions:}
    This work bridges combinatorial optimization and multi-agent planning, offering a principled framework for collaborative pathfinding with provable guarantees, and the key idea of the solution can be widely extended to many other collaborative optimization problems, such as \textsc{mapf}.
\end{abstract}



\received{20 February 2007}
\received[revised]{12 March 2009}
\received[accepted]{5 June 2009}

\maketitle

\section{Introduction}
\label{intro}
``Unity makes strength'', that small entities together can achieve great things, is a clear description of the meaning of teamwork. Thousands of robots together can handle the whole packaging process in a warehouse; hundreds of drones together can perform a breathtaking light show; millions of autonomous driving cars might replace all the traffic in a city in the near future. Promising as it sounds, how to make sure a large number of agents efficiently work together to achieve their goals is still an unsolved problem and remains an intriguing research problem. 

Multi-Agent Path Finding (\textsc{mapf}) is one such field that emphasizes teamwork of multiple agents without collision in a shared space. Solving \textsc{mapf} is a crucial problem in artificial intelligence and robotics, as it enables many real-world applications, e.g., drone swarm control~\cite{hu2021distributed}, warehouse robots~\cite{ma2014path}, and public transportation scheduling~\cite{adler2002cooperative}.  
However, collision avoidance usually separates agents in shared spaces, while there also exist scenarios where collaboration between nearby robots may be beneficial, requiring them to come close to and interact with each other when necessary.

To explore the possibility of team coordination~\cite{yan2013survey} on top of \textsc{mapf}, a new problem, Team Coordination on Graphs with Risky Edges (\textsc{tcgre}), is recently proposed~\cite{limbu2023team}, where multiple agents, under central control, travel from their starts to goals with possible support from some nodes that reduces the cost of traversing certain risky edges. In contrast to the original \textsc{mapf} that avoids collision and thus coordination among agents, \textsc{tcgre} endeavors to achieve team coordination between support nodes and risky edges to improve the overall team performance, e.g., one agent extinguishing fire on the path of another agent. The introduction of team coordination, nevertheless, complicates the problem with more possibilities. 

The first attempt to obtain optimal coordination is to construct a single-agent shortest path problem by converting the environment graph to a Joint State Graph (\textsc{jsg})~\cite{limbu2023team}. However, the conversion to \textsc{jsg} does not scale well with large environment graphs and number of robots. To address the curse of dimensionality, a Critical Joint State Graph (\textsc{cjsg}) approach has been proposed for large graphs that assures solution optimality by eliminating unnecessary edges, but it only works with a small amount of support with up to two robots. Reinforcement Learning (RL) has been utilized~\cite{limbuteam} to reduce the time complexity and scale the solution to a large group of robots and size of graph, but at the cost of sacrificing optimality. 

In this paper, we reformulate \textsc{tcgre} in a 3D Matching framework and present a rigorous mathematical analysis of this reformulated problem. We prove the NP-hardness of \textsc{tcgre} by reduction from the Minimum 3D Matching problem. We further show that such a difficult combinatorial optimization problem can be effectively addressed by efficient decomposition. In addition to providing a theoretical explanation for previous  algorithms, we reiterate them as different decomposition approaches of the 3D Matching problem: (1) The construction of \textsc{jsg} matches support pairs and robot pairs, converting it into a single-agent shortest path problem. (2) Inspired by the Conflict-Based Search~\cite{sharon2015conflict}, Coordination-Exhaustive Search (\textsc{ces}) starts with individual optimal paths and finds the lowest cost among every possible coordination for every robot to achieve the optimal solution within polynomial time with respect to the number of robots. (3) The last class of solutions deploy Receding-Horizon Optimistic Cooperative A* (\textsc{rhoc-a*}) search that assigns robot pairs with a limited horizon.
We summarize all the algorithms as different ways of decomposition in the 3DM framework. Moreover, we propose a new algorithm \textsc{Dynamic-hjsg}, which dynamically constructs a joint-state graph during graph search with limited neighborhood and incremental goal satisfaction, greatly reducing runtime and in many cases in polynomial time wrt. the number of agents, while retaining provable optimality.
\footnote{This journal paper is an extension to its conference version~\cite{zhou2024team}. Due to page limit, some details were omitted in the conference version and room for algorithmic improvement has been identified. 
This paper, as an extension, will revisit the problem, reformulate the problem as a Minimum 3D Matching problem, provide insights into existing methods, develop more efficient algorithms.}

\section{Related Work}
\label{related}
\label{sec::related}

We first review related work on the classical \textsc{mapf} problem and common classes of algorithms. We then review previous approaches to solve the \textsc{tcgre} problem. 

\subsection{\textsc{mapf} and Classes of Algorithms}

\textsc{mapf} is a specific type of multi-agent planning problem with a key constraint that no agents can collide with one another~\cite{stern2019multi}. 
A feasible solution to the problem is a joint plan that allows all agents to reach their goals from their starts. Two common objectives are makespan and total cost. Classical \textsc{mapf} problems may include additional assumptions, such as no vertex conflict, no edge conflict, no cycle conflict, and no swapping conflict~\cite{standley2010finding,felner2017search}. 

Algorithms to solve \textsc{mapf} include A*-based search with exponential space and time compelexity~\cite{ryan2008exploiting,standley2010finding}; conflict-based search~\cite{sharon2015conflict} by decomposing into many constrained single-agent problems; reduction-based approaches to SAT~\cite{surynek2012towards,surynek2016makespan}, ILP~\cite{yu2013planning}, ASP~\cite{erdem2013general}, or CSP~\cite{surynek2016efficient,bartak2017modeling}; rule-based algorithms based on Kornhauser’s algorithm~\cite{Kornhauser1984}, Push-and-Rotate~\cite{de2014push}, or BIBOX~\cite{surynek2009novel}; and suboptimal solutions~\cite{holte1996hierarchical} that sacrifice optimality for efficiency. 

\textsc{mapf} is NP-hard~\cite{goldreich2011finding}, and no optimal solutions can be found in polynomial time. The time complexity of all above optimal algorithms~\cite{ryan2008exploiting,standley2010finding,surynek2012towards,surynek2016makespan,sharon2015conflict,yu2013planning,erdem2013general,surynek2016efficient,bartak2017modeling} is  exponential to the number of agents. 
Similarly, we prove in this paper that our \textsc{tcgre} problem that utilizes, instead of avoiding, interactions between agents in the form of support is also NP-hard.

\subsection{Team Coordination on Graphs with Risky Edges (\textsc{tcgre})}
\textsc{tcgre}~\cite{limbu2023team} is a recently proposed problem, in which a team of robots traverses a graph from their starts to goals and supports each other while traversing certain risky (high-cost) edges to reduce overall cost. Instead of focusing on collision-free paths in the traditional \textsc{mapf}, the \textsc{tcgre} problem pursues team coordination. 
To solve \textsc{tcgre}, Limbu et al.~\cite{limbu2023team} have proposed \textsc{jsg} and \textsc{cjsg}, both of which construct a single-agent joint-state graph. After the construction, the original team coordination problem can be solved using Dijkstra's algorithm to solve a shortest path problem with optimality guarantee. The \textsc{cjsg} construction deals with the team coordination problem more efficiently, although it can only solve problems with two agents. 
To scale up \textsc{tcgre}, RL~\cite{limbu2023team} has been utilized to handle many nodes and robots, but at the cost of optimality.  

In this paper, we reformulate \textsc{tcgre} as a constrained optimization problem and conduct a mathematical analysis of this problem. We prove its NP-hardness and point out the necessity of efficient decomposition to effectively solve this problem. We further present three classes of algorithms to solve \textsc{tcgre} from different perspectives. 
Compared to the preliminary results reported in the conference version of this paper~\cite{zhou2024team}, we further improve the efficiency of the three classes of algorithms using homogeneity of robots and conduct comprehensive and systematic experiments to verify the near polynomial time complexity of the proposed algorithm in practice.

\section{Problem}
\label{problem}
Assuming a team of $N$ homogeneous robots traverses an undirected graph \(\mathbb{G}=(\nodeset,\edgeset)\), where \(\nodeset\) is the set of nodes the robots can traverse to and \(\edgeset\) is the set of edges connecting the nodes, i.e., \(\edgeset\subset \nodeset\times \nodeset\). 
The team of robots traverses in the graph from their start nodes $\nodeset_0 \subset \nodeset$ to goal nodes $\nodeset_g \subset \nodeset$ via edges in \(\edgeset\). Each edge $e_{ij}=(V_i,V_j)\in \edgeset$ is associated with a cost $c_{ij}$, depending on its length, condition, traffic, obstacles, etc. Specially, some edges with high costs are difficult to traverse through, denoted as risky edges $\edgeset' \subset \edgeset$, but with the support from a teammate from a supporting node, their costs can be significantly reduced to $\tilde{c}_{ij}$. 
In this problem, we only consider such coordination behaviors between two robots. In one coordination behavior, one receiving robot receives support while traversing a risky edge, and another supporting robot offers support from some (nearby) location, called support node. Note that each risky edge $e_{ij} \in \edgeset'$ corresponds to certain support node(s) $\mathbb{S}_{e_{ij}} \subset \nodeset$ ($\mathbb{S}_{e_{ij}}=\emptyset$ if $e_{ij}\notin \edgeset'$). Denote the set of all support nodes by $\mathbb{S}_{e}= \bigcup_{e_{ij}}\mathbb{S}_{e_{ij}}$. Additionally, the coordination also induces some cost for the supporter, denoted by $c'$.
A central planner needs to schedule the paths of all agents and coordination on their ways.

\subsection{Action \& Cost Model}
\label{CostModel}

Without coordination,
at each time step $t$, a robot $n$ can choose to stay where it is, or move to its neighbor ($V_i$ is the neighbor of $V_j$ if $e_{ij}\in \edgeset$). Its movement can be denoted by $M_n^t=(l_n^t,l_{n}^{t+1}) \in \edgeset$, where $l_n^t, l_{n}^{t+1} \in \nodeset$ indicate its current and next location and $l_{n}^{t+1}$ is a neighbor of $l_n^t$. Specially, the robot stays at its current location if $l_{n}^{t+1}=l_n^t$ with zero cost, i.e., $c_{ii}=0,\forall i$. 
The movement set can thus be denoted by $\mathcal{M} = \{M_{n}^{t}|\forall n, \forall t\}$. 
Moreover, the movement decision $M_{n}^{t}$ can be rewritten as an 0/1 variable $M_{ij}^{nt}$, where $M_{ij}^{nt}=1$ represents edge $e_{ij}$ is selected by robot $n$ at time $t$, and 0 otherwise. A robot can only move once at each time step to a neighbor node or not at all, i.e., $\sum_{\forall e_{ij}\in \mathcal{N}_{l_n^t}}M_{ij}^{nt}=1 ~\&~ \sum_{\forall e_{ij}\notin \mathcal{N}_{l_n^t}}M_{ij}^{nt}=0$, where $\mathcal{N}_{l_n^t}=\{(l_n^t,l_n^{t+1})|\forall (l_n^t,l_n^{t+1})\in \edgeset\}$. The movement set can thus be denoted by $\mathcal{M} = \{M_{ij}^{nt}| \forall i,j,\forall n, \forall t\}$. 

When a coordination behavior is available---when robot $n$ is going to traverse a risky edge, another robot $m$ happens to be at one of the support nodes of the risky edge or vice versa, i.e., $M_n^t\in \edgeset'$ and $l_m^t\in \mathbb{S}_{M_n^t}$, or $M_m^t\in \edgeset'$ and $l_n^t\in \mathbb{S}_{M_m^t}$---the robot pair needs to decide whether to provide/receive support. 
Denote the coordination decision of agent $n$ at time $t$ as $s_{nm}^t$. It is clear that agent $n$'s coordination decision is dependent on its movement decision, so the cost is twofold:
\noindent (1) When agent $n$ has no coordination opportunity (the above coordination behavior is not available for any other robot $m$), i.e., $\forall m$, $l_m^t\notin \mathbb{S}_{M_n^t}$ and $l_n^t\notin \mathbb{S}_{M_m^t}$, its cost $C_n^t$ is only decided by its movement, i.e., $C_n^t = c_{ij}$, where $M_{ij}^{nt} = 1$. 
 
\noindent (2) When coordination is possible for agent $n$, i.e., $\exists m$, $l_m^t\in \mathbb{S}_{M_n^t}$ or $l_n^t\in \mathbb{S}_{M_m^t}$, the cost $C_n^t$ can be represented as
\begin{align}
\label{IndividualCost}
    C_n^t=\begin{cases}
        c_{ij}, ~~\textit{if} ~~ s_{nm}^t=~~0;\\
        \tilde{c}_{ij}~~~\textit{if} ~~ s_{nm}^t=~~1;\\
        c',~~~\textit{if} ~~ s_{nm}^t=-1.
    \end{cases}
\end{align}
\noindent where $s_{nm}^t=1$ means agent $n$ decides to receive support from $m$,  $s_{nm}^t=-1$ indicates agent $n$ decides to offer support to $m$, and $s_{nm}^t=0$ stands for no coordination between the robot pair $n$ and $m$ at $t$.
Specially, no coordination happens for one single robot, i.e., $s_{nn}^t=0,\forall n,\forall t$, or when $n$ and $m$ cannot support each other, i.e., $s_{nm}^t=0$ if $l_m^t\notin \mathbb{S}_{M_n^t}$ and $l_n^t\notin \mathbb{S}_{M_m^t}$, $\forall m$. 
The coordination decision set can be written as $\mathcal{S}=\{s_{nm}^t|\forall n,m, \forall t\}$. In addition, a coordination decision is made for a pair, so $s_{nm}^t+s_{mn}^t=0$ for every pair of robots. Furthermore, the robots can wait now (no movement) for future coordination, but there is no point for all robots to stay still at the same time, i.e.,  $\sum_{\forall n}\sum_{\forall i\neq j} M_{ij}^{nt}\neq0$.

\subsection{Problem Definition}
\label{definition}
Given the node set $\nodeset$, the edge set $\edgeset$, support nodes for each edge $\mathbb{S}_{e_{ij}}$, cost of each edge without and with coordination $c_{ij},\tilde{c}_{ij}$, $N$ robots with their starts $\nodeset_0$ and goals $ \nodeset_g$, optimize the movement and coordination decisions $\mathcal{M}$ and $\mathcal{S}$, in order to minimize the total cost for each agent to traverse from its start to its goal within a time limit $T$.
Formally, the problem can be represented as
\begin{gather}
    \min_{{\mathcal{M}},{\mathcal{S}}} \sum_{t=0}^{T-1} \sum_{n=1}^N C_{n}^t. \label{objective}\\
    \allowdisplaybreaks[4]
    s.t.~~
     \sum_{\forall m \in \{1,2,...,N\}}|s_{nm}^t|\leq 1,\nonumber\\
     \allowdisplaybreaks[4]
    \forall n\in \{1,2,...,N\}, \forall t\in \{0,1,...,T-1\}. \label{constraint1:supportnum}\\
    \allowdisplaybreaks[4]
    s_{nm}^t, s_{mn}^t\in\{-1, 0,1\}, s_{nm}^t+ s_{mn}^t=0, \nonumber\\
    \allowdisplaybreaks[4]
    \forall n,m\in \{1,2,...,N\}, \forall t\in \{0,1,...,T-1\}. \label{constraint2:supportpair}\\
    \allowdisplaybreaks[4]
    l_n^0=\nodeset_0(n), l_n^T=\nodeset_g(n),\forall n\in \{1,2,...,N\}.\label{constraint3:startsandgoals}\\
    \allowdisplaybreaks[4]
    \sum_{\forall e_{ij}\in{\mathcal{N}_{l_n^t}}}M_{ij}^{nt}=1,  \sum_{\forall e_{ij}\notin{\mathcal{N}_{l_n^t}}}M_{ij}^{nt}=0,                   \nonumber\\
    \allowdisplaybreaks[4]
    \forall n\in \{1,2,...,N\}, \forall t\in \{0,1,...,T-1\}.\label{constraint5:movenum}\\
    \allowdisplaybreaks[4]
    \sum_{\forall n}\sum_{\forall i\neq j} M_{ij}^{nt}\neq0,\nonumber\\
    \forall t\in\{0,1,...,T-1\}.\label{constraint6:correlation}
\end{gather}

\noindent 
Eqn.~\eqref{objective} suggests the goal of the problem is to minimize the total cost of all agents across all time steps with two decision variables, movement set $\mathcal{M}$ and coordination set $\mathcal{S}$. Eqn.~\eqref{constraint1:supportnum} indicates that, at each time step, each robot can participate in at most one coordination behavior. Eqn.~\eqref{constraint2:supportpair} regulates that, at each time step, one coordination behavior only occurs between one robot pair. Eqn.~\eqref{constraint3:startsandgoals} sets the start and the goal for each robot. Eqn.~\eqref{constraint5:movenum} guarantees that, at each time step, a robot can only move to a neighbor node or stay still. Eqn.~\eqref{constraint6:correlation} assures no unnecessary stagnation.

\section{Problem Analysis \&  Reformulation}
\label{analysis}
In this section, we first conduct a full analysis on the problem and explain why we can reformulate it as a matching problem. Next, we prove our \textsc{tcgre} problem reduces from the Maximum 3D Matching problem, an NP-hard problem. Last, we provides the 3DM model of the problem.

\subsection{Problem Analysis}
\label{Analysis}
We can reformulate the problem in the 3D Matching framework with robot pairs, support pairs, and time orders. 
\subsubsection{Time Order with A Simplified Graph}
\begin{figure}
    \centering
    \includegraphics[width=0.8\linewidth]{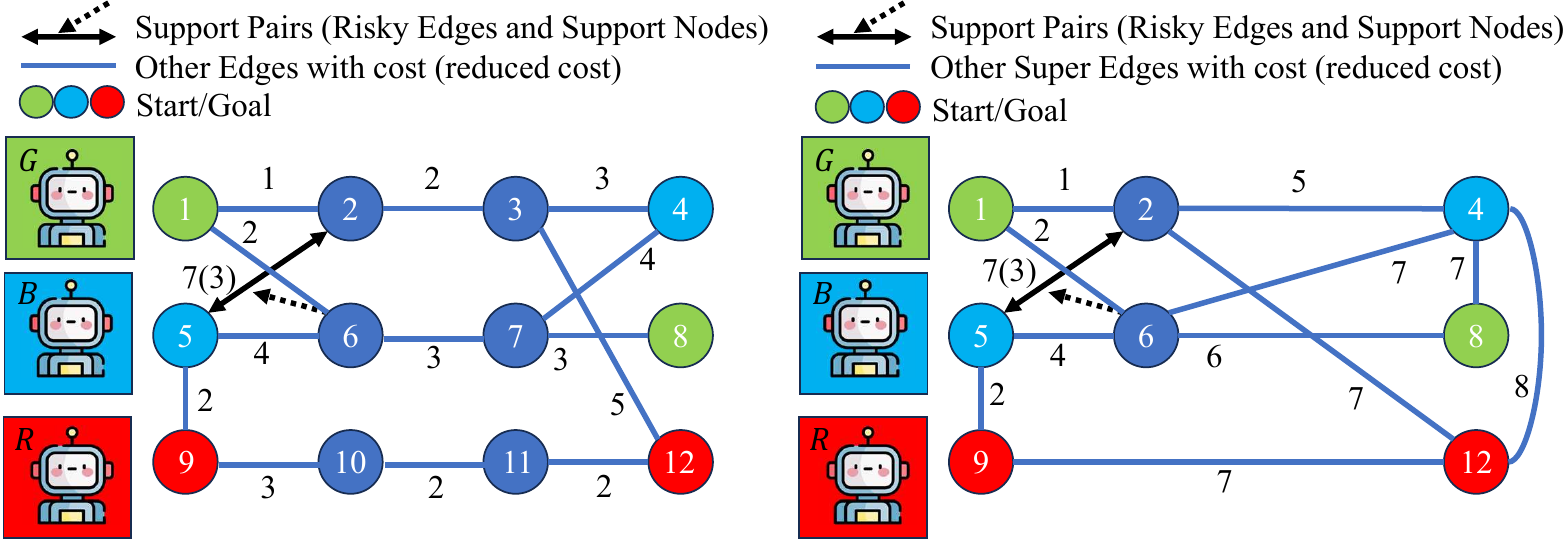}
    \caption{Left: original graph $\mathbb{G}$; right: simplified graph $\mathbb{G}_s$ with fewer nodes.}
    \label{simplfied_graph}
\end{figure}

\
\label{simplifiedgraph}
Time steps don't matter because we allow robots to stay still for future coordination. What matters is the time order of each coordination behavior. Furthermore, interaction among robots only happen when they are at some specific nodes, i.e., endpoints of risky edges and support nodes. In other words, robots' movement involving other nodes is independent from the collaborative optimization and simply many single-agent shortest path problems. Therefore, we could simplify the original graph with only these special nodes plus the start and end nodes of all robots, i.e.,$\nodeset_s = \{V_i, V_j| \forall e_{ij}\in \edgeset'\} \cup \mathbb{S}_e \cup \nodeset_0 \cup \nodeset_g$. Then, the edge between each two nodes is either the original shortest path between them without coordination and with the edge cost being the cost of the path, or just a risky edge with original edge cost or reduced cost depending on coordination. Denote the edge set as $\edgeset_s$. Now, the original graph can be equivalently converted to a simplified graph $\mathbb{G}_s = (\nodeset_s, \edgeset_s)$, as in Fig.~\ref{simplfied_graph}. (Some super edges are pruned as they involve intermediate special nodes, e.g., the edge between node 6 and node 9 (right) because it passes through node 5 (left). This means we only save super edges without passing through special nodes.)
To distinguish the two graphs, we call a node in the simplified graph $\mathbb{G}_s$ a super node $Vs_i$, and an edge a super edge $es_{ij}=(Vs_i,Vs_j)$ with edge cost 
\begin{align}
\label{SuperEdgeCost}
    cs_{ij}=\begin{cases}
        spc(Vs_i,Vs_j), ~~~~~~~~~~~~~~\textit{if} ~~ es_{ij}\notin\edgeset';\\
        \min(c_{ij},spc(Vs_i,Vs_j)),~~\textit{if} ~~ es_{ij}\in\edgeset'~\land~s_{nm}^t=0;\\
        \min(\tilde{c}_{ij},spc(Vs_i,Vs_j)),~~\textit{if} ~~ es_{ij}\in\edgeset'~\land~s_{nm}^t=-1,
    \end{cases}
\end{align}
where $spc()$ is the shortest path cost between two nodes in $\mathbb{G}$. Note that the superscript $t$ means the time order of events (with possible one or multiple time steps) instead of one time step, here and in the rest of the paper. Plus, a robot cannot support others while traversing a risky edge, so $s_{nm}^t=1$ is impossible here. Moreover, coordination is meaningless if the reduced cost of a risky edge is even larger than the shortest path cost between these two nodes, in which case we would simply regard this edge as a normal edge, remove it from the risky edge set $\edgeset'$ and its endpoints from the special nodes, and simplify the graph further. Afterwards, we can ensure coordination between a robot pair will reduce the cost of a certain super edge. 

\subsubsection{Robot Pair Cost}
\label{robotpaircost}
Because we only care about the total cost of all robots, and every coordination requires a robot pair, we can reassign the coordination cost $c'$ from the supporter to the receiver in addition to the reduced edge cost $\tilde{c}_{ij}$, without changing the problem.  
So, the cost with coordination in Eqn.~\eqref{IndividualCost} becomes

\begin{align}
\label{NewIndividualCost}
    C_n^t=\begin{cases}
        cs_{ij}, ~~\textit{if} ~~ s_{nm}^t=~~0;\\
        \hat{cs}_{ij},~~\textit{if} ~~ s_{nm}^t=~~1;\\
        0,~~~~~\textit{if} ~~ s_{nm}^t=-1.
    \end{cases}
\end{align} 
where $\hat{cs}_{ij}=\tilde{cs}_{ij}+c'$ and $\tilde{cs}_{ij}$ is the third condition $cs_{ij}$ in Eqn.\eqref{SuperEdgeCost}. 

The cost of a robot pair $C_{nm}^t$ is the sum of both $C_{n}^t$ and $C_{m}^t$
\begin{align}
\label{RobotPairCost}
    C_{nm}^t=\begin{cases}
        cs_{i_n j_n}+cs_{i_m j_m}, ~~\textit{if} ~~ s_{nm}^t=~~0 ~\land~ s_{mn}^t=~~0;\\
        \hat{cs}_{i_n j_n},~~~~~~~~~~~~~~\textit{if} ~~ s_{nm}^t=~~1~\land~s_{mn}^t=-1;\\
        \hat{cs}_{i_m j_m},~~~~~~~~~~~~~\textit{if} ~~ s_{nm}^t=-1~\land~s_{mn}^t=~~1.
    \end{cases}
\end{align}
Obviously, when robots $n$ and $m$ don't coordinate ($s_{nm}^t=0$), the robot pair cost is the sum of the original costs of both edges they traverse. Otherwise, it is $\hat{c}_{ij}$, depending on which robot is receiving support while traversing a risky edge. 
Therefore, the original objective function (Eqn.~\eqref{objective}) can be rewritten as
\begin{gather}
\min_{{\mathcal{M}},{\mathcal{S}}} \sum_{t=0}^{T-1} \sum_{n\neq m} \sum_{\forall es_{ij}}M_{ij}^{nt}[(1-s_{nm}^t)c_{ij}+s_{nm}^t \hat{c}_{ij}] + M_{ij}^{mt}[(1-s_{mn}^t)c_{ij}+s_{mn}^t \hat{c}_{ij}]\label{new_objective},\\
    s.t. ~~
    \eqref{constraint1:supportnum}, \eqref{constraint2:supportpair}, \eqref{constraint3:startsandgoals}, 
    \eqref{constraint5:movenum}, \eqref{constraint6:correlation}.\nonumber
\end{gather}
\noindent Notice that because when $s_{nm}^t=-1$, $M_{ij}^{nt}=0,\forall i\neq j$, the last \textit{if} condition in Eqn.~\eqref{NewIndividualCost} does not need to be considered in Eqn.~\eqref{new_objective}.  The three summations indicate time order, robot pairs, and super edges. 

\subsubsection{Support Pair based Matching}
\label{supportpairmatching}
Now, we could construct a matching problem, by extracting support pairs (one support pair refers to one risky edge and one of its support nodes). Denote a support pair as $p_{ijk}=(es_{ij},es_{kk})$ where $es_{ij}\in \edgeset'~\&~ Vs_k\in\mathbb{S}_{es_{ij}}$, and the set of support pair as $\mathbb{P}=\{p_{ijk}|\forall es_{ij}\in \edgeset' ~\&~ \forall Vs_k\in\mathbb{S}_{es_{ij}}\}$.
As shown in Eqn.~\eqref{SuperEdgeCost}, the cost of a normal super edge is a constant value and only the risky edges involve coordination decisions between robot pairs.  Therefore, we can further convert the objective function as 
\begin{gather}
    \min_{{\mathcal{M}},{\mathcal{S}}} 
    \overbrace{\sum_{t=0}^{T-1}  \sum_{\forall n,m}\sum_{\forall es_{ij}} (M_{ij}^{nt}+M_{ij}^{mt})cs_{ij}}^{\textrm{Cost without Coordination}} -
    \underbrace{\sum_{t=0}^{T-1} \sum_{\forall n,m} 
    \sum_{\forall p_{ijk}}
    (M_{ij}^{nt}-M_{ij}^{mt})s_{nm}^t\Delta{c}_{ij}}_{\textrm{Cost Reduction due to Coordination}}, \label{coordinationbasedobjective1}
\end{gather}
where $\Delta{cs}_{ij} = {c}_{ij}-\hat{cs}_{ij}$ and either $M_{kk}^{nt}=1$ or $M_{kk}^{mt}=1$. When $s_{nm}^t = 1$ (n receiving support) the second part reduces the cost in the first part by $\Delta{cs}_{ij}$, based on which risky edge, $es_{ij}$, $n$ is traversing; When $s_{nm}^t = -1$ (n providing support), the cost reduction depends on which risky edge $m$ is traversing.

Note that the first half of Eqn.~\eqref{coordinationbasedobjective1} is the summation of all super edges, which includes risky edges similar to the second half, and non-risky edges. As mentioned in Sec.~\ref{simplifiedgraph}, super edges that involve no coordination can be seen as independent piecewise shortest path problems which we already solved as the first condition in Eqn.~\eqref{SuperEdgeCost}. That is to say, the first half can be seen as a function of matching weights among time order, robot pairs, and support pairs, similar to the second half, additionally with some support pair-related parameters $spc_{ij}$ (piecewise shortest path cost):
\begin{gather}
    \min_{{\mathcal{M}},{\mathcal{S}}} 
    \overbrace{\sum_{t=0}^{T-1}  \sum_{\forall n,m}\sum_{\forall p_{ijk}} (M_{ij}^{nt}+M_{ij}^{mt})(cs_{ij}+spc_{ij}) }^{\textrm{Cost without Coordination}} -
    \underbrace{\sum_{t=0}^{T-1} \sum_{\forall n,m} 
    \sum_{\forall p_{ijk}}
    (M_{ij}^{nt}-M_{ij}^{mt})s_{nm}^t\Delta{c}_{ij}}_{\textrm{Cost Reduction due to Coordination}}, \label{coordinationbasedobjective2}
\end{gather}
Now, we can reformulate the problem as a minimum 3D Matching Problem, as the summations are on time orders, robot pairs, and support pairs.

\subsubsection{NP-Hardness}
\label{npharness}
\begin{definition}
\label{3DMatching}
    Minimum 3D Matching (\textsc{minimum 3dm}): ${\bf{X}},{\bf{Y}},{\bf{Z}}$ are 3 finite sets. ${\bf{T}}$ is the subset of ${\bf{X}}\times{\bf{Y}}\times{\bf{Z}}$, with triples $(x,y,z)$, where $x\in {\bf{X}},y\in{\bf{Y}},z\in{\bf{Z}}$. ${\bf{M}}\subset  {\bf{T}}$ is a 3D matching if for any two distinct triples $(x_1, y_1, z_1)$ and $(x_2, y_2, z_2) \in {\bf{M}}$, we have $x_1 \neq x_2$, $y_1 \neq y_2$, and $z_1 \neq z_2$; each triple has a weight $w(x_i,y_j,z_k)$. Minimum 3D Matching problem is 
    to find a 3D matching with minimum total weight.
\end{definition}

\begin{figure}[ht]
    \begin{minipage}[t]{0.3\linewidth}
    \centering
    \raisebox{0.13\height}{\includegraphics[width=0.99 \textwidth]{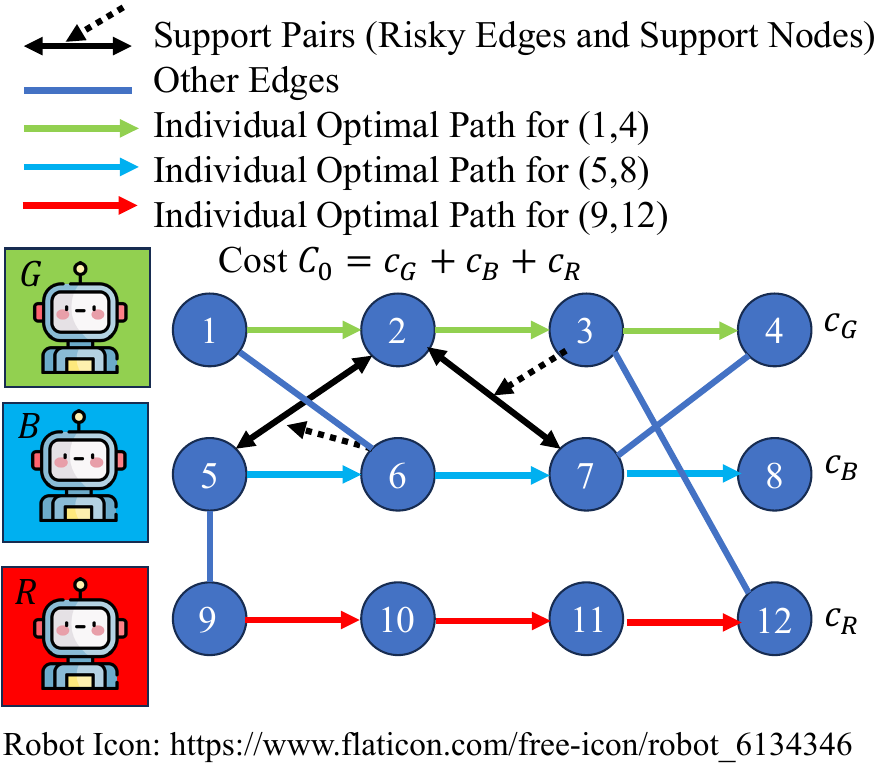}}
    \label{IndividualPaths}
    \end{minipage}
    \begin{minipage}[t]{0.38\linewidth}
    \centering
    \includegraphics[width=0.99\textwidth]{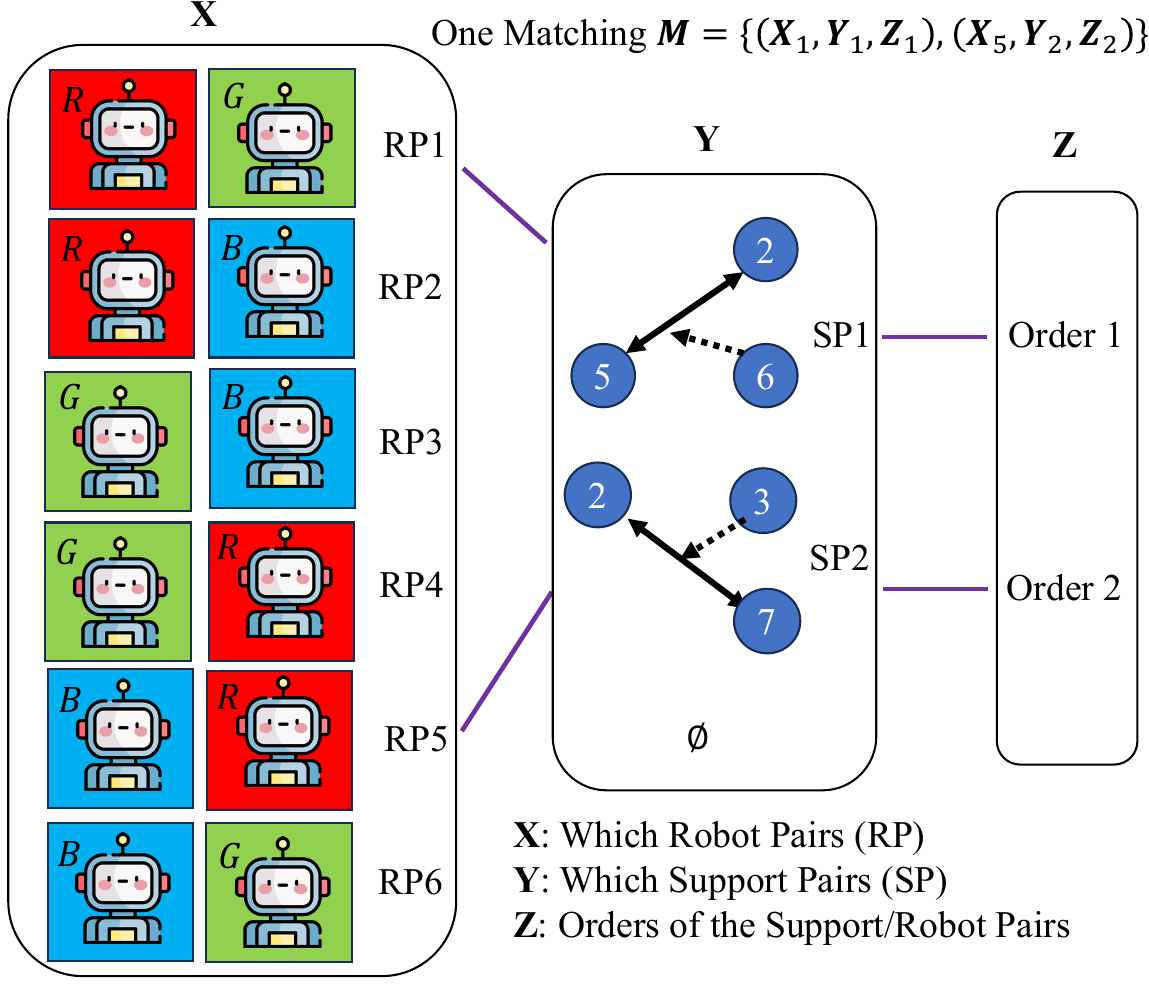}
    \label{3DMatchingGraph}
    \end{minipage}
    \begin{minipage}[t]{0.3\linewidth}
    \centering
    \raisebox{0.22\height}{\includegraphics[width=0.99 \textwidth]{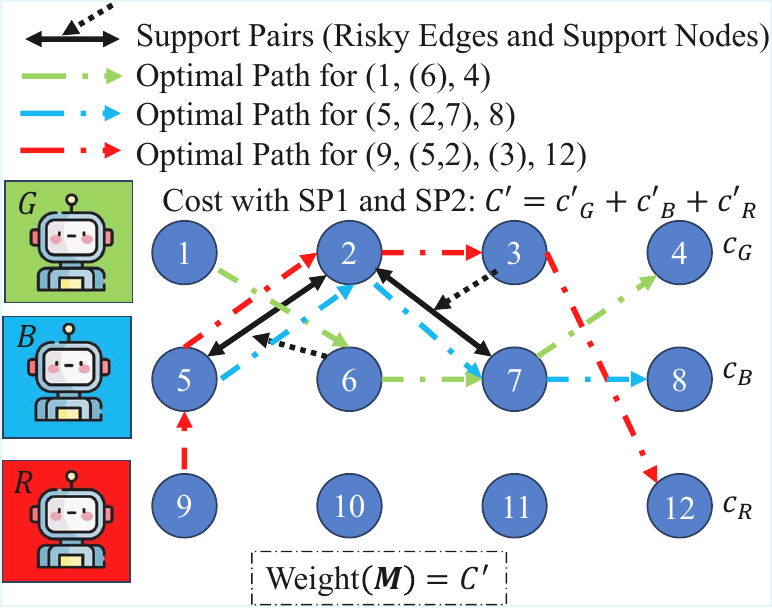}}
    \label{ReducedCost}
    \end{minipage}

\caption{Reduction from Minimum 3D Matching (Middle) to \textsc{tcgre} (Left and Right). }
\label{Proof1}

\end{figure}
\begin{theorem}
\label{shortestpathand2dmatching}
    \textsc{tcgre} reduces from \textsc{minimum 3dm} of robot pairs, support pairs, and time steps with transition-dependent costs.
\end{theorem}
\begin{proof}
Consider {\bf{X}} contains all robot pairs, i.e., ${\bf{X}}=\{(n,m)| n\neq m, \forall n,m\in\{1,2,...,N\}\}$; ${\bf{Y}}$ is the set of all support pairs plus an empty set element, i.e., ${\bf{Y}}=\mathbb{P}\cup\{\emptyset\}$; {\bf{Z}} is a list of time orders of events (time steps are not necessarily needed, because robots can stay still and wait, Fig.~\ref{Proof1} middle). 
Consider the weight $w(x_i,y_j,z_k)$ as the cost of after-coordination path cost for a robot pair $x_i$ to detour to support pair $y_j$ with time order $z_k$. The weight of a 3D matching $\textbf{M}$ is the total cost of assigning all robot pairs to one support pair. Considering the constraint that, in one matching, each support pair can be assigned at most once ($y_1 \neq y_2$ as in the definition), to make the reconstructed matching problem equivalent as the original formulation, we additionally expand ${\bf{Y}}$ to ${\bf{Y}}'$, by duplicating all non-empty set elements for a large number of time $D$, i.e., ${\bf{Y}}'=\{\emptyset\}\cup \bigcup_{d=1}^D \mathbb{P}$. Similarly, each robot pair might be assigned to all support pairs, so we need to duplicate all robot pairs by the number of support pairs, i.e., ${\bf{X}}'=\bigcup_{d=1}^{|{\bf{Y}}'|}{\bf{X}}$, which is related to $D$ as well.
Therefore, \textsc{tcgre} with the constraint that each support pair can be only assigned at most $D$ times is Transition-Dependent 3D Matching (\textsc{td-3dm}) of ${\bf{X}}', {\bf{Y}}',$ and ${\bf{Z}}$. The constraint only exists because in 3D Matching, the size of a set has to be finite; otherwise, $D=\infty$ as in the original \textsc{tcgre}. Furthermore, since there are no negative edge costs in our graph, in any feasible solution $D$ will not be infinite. That is to say, the constraint doesn't change complexity of \textsc{tcgre}; without the constraint, it will still be as hard as \textsc{td-3dm}, which is already a generalized form of 3DM with the weight of a triple dependent on other triples in the matching ($w(x_{1},y_{1},z_{1})$ depends on $w(x_{2},y_{2},z_{2})$ if $z_{2}<z_{1}$), i.e., $\textsc{minimum 3dm}\leq_P \textsc{minimum td-3dm}\leq_P \textsc{tcgre}$.

Since \textsc{tcgre} reduces from \textsc{minimum 3dm}, a classical NP-hard problem~\cite{kann1991maximum}, \textsc{tcgre} is also NP-hard. We cannot find an optimal solution in polynomial time.    
\end{proof}


\subsection{\textsc{tcgre} as \textsc{Minimum 3dm}}
\label{TCGREas3DM}
\textsc{tcgre} can be reformulated in the following form\footnote{The subscripts $i,j,k$ in the reconstructed problem (from Sec.~\ref{npharness} to Sec.\ref{TCGREas3DM}) have different meanings from other sections.}: \\
Given three disjoint sets of robot pairs ${\bf{X}}'=\bigcup_{d=1}^{|{\bf{Y}}'|}\{(n,m)| n\neq m, \forall n,m\in\{1,2,...,N\}\}$, support pairs ${\bf{Y}}'=\{\emptyset\}\cup \bigcup_{d=1}^D \mathbb{P}$, and time orders ${\bf{Z}}= \{0,1,...,T-1\}$, where ${\bf{T}}$ is the subset of ${\bf{X}}'\times {\bf{Y}}'\times {\bf{Z}}$, each triple with weight $w_{ijk}$, we need to find a matching $\mathbf{M}$ --- a special subset $\mathbf{M}\subset{\mathbf{T}}$ where every element is disjoint in each dimension, i.e., $\forall (x_1, y_1, z_1), (x_2, y_2, z_2) \in {\bf{M}}$, $x_1 \neq x_2$ \&  $y_1 \neq y_2$ \& $z_1 \neq z_2$  --- such that the total weight of the matching is minimum. Let $a_{ijk}$ denote if the triple $(x_i,y_j,z_k)$ is selected in the matching, i.e., $a_{ijk}=1$ if $(x_i,y_j,z_k)\in \mathbf{M}$, and $0$ otherwise. Formally,
\begin{gather}
    \min_{{\bf{M}}\subset{{\bf{T}}}} \sum_{(x_i,y_j,z_k)\in\mathbf{T}} a_{ijk}\cdot w_{ijk}. \label{3DMobjective}\\
    s.t. \sum_{jk:(x_i,y_j,z_k)\in\mathbf{T}} a_{ijk}\leq 1, ~\forall x_i\in {\bf{X}}'.\label{3DMconstraint1}\\
    ~~~~~\sum_{ik:(x_i,y_j,z_k)\in\mathbf{T}} a_{ijk}\leq 1, ~\forall y_j\in {\bf{Y}}'.\label{3DMconstraint2}\\
    ~~~~~\sum_{ij:(x_i,y_j,z_k)\in\mathbf{T}} a_{ijk}\leq 1, ~\forall z_k\in {\bf{Z}}.\label{3DMconstraint3}
\end{gather}
Note that the set of support pairs ${\bf{Y}}'$ is constructed based on the graph information, detailed in Sec.~\ref{supportpairmatching}.
The three constraints together guarantee that every element in the matching is disjoint in each dimension. (This problem could also be formed as 3DM of super nodes, robots, and time orders, but with additional constraints of coordination behaviors, which looks closer to a general ILP problem.)
However, different from traditional 3DM, the weight of a triple $w_{ijk}$ (as in Eqn.~\eqref{coordinationbasedobjective2}) in our problem is dependent on other triples prior in terms of the time order, which is a more generalized problem, we call Transition-Dependent 3DM (\textsc{td-3dm}). Similar to any collaborative optimization problem, breaking down the problem in a way will make the solution more efficient.




\section{Methods}
\label{methods}
In the conference version~\cite{zhou2024team}, we proposed 3 classes of algorithms, \textsc{jsg}, \textsc{ces}, and \textsc{rhoca}, which we later found were "decomposition" of the 3DM problem into two 2DM problems in different ways.
In this section, we briefly recapitulate these algorithms, under the 3DM framework. Then, we propose an algorithm that merges characteristics of all the algorithms, called Homogeneous Joint-State Graph (\textsc{hjsg}), as it utilizes homogeneity of the robots and to distinguish it from the original \textsc{cjsg}~\cite{limbu2023team}.

\subsection{Algorithm Recap}
Intuitively, we could simulate 3DM with two 2DMs, however, with some interdependency, which becomes more complex when the weights are transition dependent. That is to say, we can't treat it as two independent 2DM problems, unless we sacrifice optimality, but this "decomposition" process is still helpful in solving collaborative optimization. 

\textsc{jsg} constructs joint-states and convert the problem into a single-agent shortest path problem, which is a transition-dependent 2DM problem between joint-states and time steps. The construction of all joint-states embeds all possible matchings between robot pairs and support pairs. In short, implicitly, \textsc{jsg} first matches robot pairs and support pairs, and then time steps.

\textsc{ces} tries to directly solve it as 3DM, with 3 loops that iterate through all possible matchings. More specifically, it first checks every order of support pairs, and then assigns at most one robot pair for each support pair, which means \textsc{ces} first matches time orders and support pairs, and then robot pairs.

\textsc{rhoca} follows a simple idea --- at each time step, pick a robot pair to move within a limited horizon. If we apply it to the simplified graph, it would be first matching time orders with robot pairs, and then support pairs.

We now realize it is not a coincidence that we came up with 3 classes of solutions; though the algorithm details still have room for improvement, in terms of the general process, they are the only three "decompositions" of the 3DM problem. Based on the fact that, in our problem, the weight of a triple is dependent on transition, it is the best that we match time order last to preserve optimality, which means \textsc{jsg} deserves more exploration.

\subsection{\textsc{Dynamic-hjsg}}
\label{hjsg}
\textsc{jsg}-based solutions perform the decomposition into the two sub-problems by 1) implicitly solving $\mathcal{S}$ by calculating the minimum edge cost for each edge in the \textsc{jsg}; 2) explicitly solving $\mathcal{M}$ by solving a single-robot shortest path problem with $\mathcal{S}$ implicitly encoded. 
Similar to previous method (\textsc{cjsg}), \textsc{hjsg} also focuses on the critical nodes and edges in this problem that involve possible coordination behaviors. In this section, we first introduce the general steps of the algorithm, and then explain algorithm details that greatly improves the efficiency.

\subsubsection{Basic HJSG Construction}
\label{HJSG}
In a joint-state graph (JSG), one joint-state is the set of all robots' locations ${\mathbf{JS}}=\{l_1,l_2,...,l_N\}$. Here we build the joint-state graph on top of the simplified graph $\mathbb{G}_s$, meaning the location of each robot  must be one super node in the simplified graph as defined in Sec.~\ref{simplifiedgraph}, i.e., $l_n\in \nodeset_s$. Denote the set of all joint-states as $\mathbb{JS}$. Next, we define joint-edges between joint-states, along with their associated costs. 
A joint-edge exists between two joint-states ${\mathbf{JS}}_i$ and ${\mathbf{JS}}_j$ if, for every robot $n$, there exists a super edge connecting $l^n_i$ and $l^n_j$, where $l^n_i \in {\mathbf{JS}}_i$ and $l^n_j \in {\mathbf{JS}}_j$. Formally, the joint-edge is defined as ${\mathbf{JE}}_{ij}=\{(l^n_i,l^n_j)|\forall n\}$, and the set of all such joint-edges is denoted $\mathbb{JE}$.
Calculating the cost of a joint-edge is less straightforward, because 1) a risky edge can be associated with multiple support nodes, 2) a single support node may support multiple risky edges, and 3) coordination is regulated such that each coordination behavior involves exactly one support pair. Since the simplified graph ensures that coordination decreases the cost of a risky edge, we only need to consider matching risky edges and support nodes that appear within the two joint-states.
\begin{theorem}
    Matching risky edges and support nodes between two joint-states is a Maximum Weighted Bipartite Matching problem.
\end{theorem}
\begin{proof}
    Given all robots' current locations and next locations,  ${\mathbf{JS}}_i=\{l^1_i,l^2_i,...,l^N_i\}$ and ${\mathbf{JS}}_{j}=\{l^1_{j},l^2_{j},...,l^N_{j}\}$, we can build a bipartite graph with two groups of robots: 1) ${\bf{X}}=\{n~|~l^n_i=l^n_{j}\in\mathbb{S}_e\}$: robots that remain on support nodes; 2) ${\bf{Y}}=\{n~|~(l^n_i,l^n_{j})\in\edgeset'\}$: robots traversing risky edges. Other robots are not relevant in coordination. For each potential support pair, we add an edge in the bipartite graph with a weight equal to the cost reduction due to coordination. The goal is to find a matching that maximizes the total cost reduction—this is precisely the Maximum Weighted Bipartite Matching problem.
\end{proof}
\noindent Notice that this step matches support pairs and robot pairs for one joint-edge (the duplicates of them spread out to all joint-edges), and is solvable by the Hungarian algorithm~\cite{kuhn1955hungarian} in $O(N^3)$, leading to an overall time complexity of $O(N^3 \cdot |\mathbb{JE}|)$ for all joint-edges. After computing the maximum matching, we apply the reduced cost to matched risky edges and original costs to all other super edges. Their total cost becomes the joint-edge cost.

Once the JSG is constructed, the overall problem becomes a single-agent shortest path problem from the start joint-state $\nodeset_0$ to the goal joint-state $\nodeset_g$. This can be solved using Dijkstra’s algorithm~\cite{dijkstra1959note} in $O(|\mathbb{JE}|\cdot \log(|\mathbb{JS}|))$. The complexity depends heavily on the size of the state space: both $|\mathbb{JS}|$ and $|\mathbb{JE}|$ are in $O(|\nodeset_s|^N)$.

\subsubsection{Dynamic Homogeneous Joint-State Graph Search}
The major computational bottleneck in the above approach is the size of the full JSG. However, constructing the entire graph is not necessary. 
Instead, we can dynamically build the joint-state graph on-the-fly during path search. This approach reduces the number of joint-states considered by Dijkstra's algorithm. 
Moreover, since coordination occurs between robot pairs and individual robots can remain stationary, we can further limit the neighborhood of each joint-state. Specifically, we restrict transitions between joint-states such that at most two robots are allowed to move between any pair of neighboring joint-states. Formally, for a joint-edge ${\mathbf{JE}}_{ij}$, we require ${\mathbf{JE_{ij}}}=\{(l^n_i,l^n_j)\in\edgeset_s, (l^m_i,l^m_j)\in\edgeset_s|\forall n,m\}$. This constraint effectively prunes the number of neighboring joint-states and consequently reduces the number of joint-edges to evaluate, while retaining optimality. 
\begin{lemma}[Sufficiency of Local Joint-State Transitions]
\label{lemma1}
    It is sufficient to restrict joint-edge construction in the dynamic HJSG to transitions where at most two robots change locations between joint-states. That is, ${\mathbf{JE_{ij}}}=\{(l^n_i,l^n_j), (l^m_i,l^m_j)|\forall n,m\}$. Under this restriction, the dynamically constructed HJSG still contains the optimal joint-state path from start to goal.
\end{lemma}
\begin{proof}
Since coordination is defined as a cost reduction that arises only when two robots simultaneously traverse risky edges while the other occupies corresponding support nodes, any coordination opportunity in the full HJSG can be captured by allowing up to two robots to move simultaneously.

If a transition in the full JSG involves more than two robots moving at once, it can be decomposed into a sequence of transitions where at most two robots move at a time, without losing coordination opportunities or increasing the overall path cost. This is because:

1) Coordination gain is pairwise and independent across robot pairs.

2) Robots not involved in coordination can remain stationary, and their movement can be scheduled before or after the coordinated step.

Thus, the optimal joint-state path in the full JSG can be replicated using only joint-edges involving at most two robot moves per transition. As Dijkstra's algorithm explores all reachable nodes under these transitions, the dynamically constructed HJSG suffices to find the optimal path.
\end{proof}

After this step, each joint-state only expands to a neighboring joint-state with at most two agents moving, which means there exists at most one support pair and one possible cost reduction for a risky edge between two neighbors. Plus, in the simplified graph we guarantee the cost of a risky edge is always reduced if coordination occurs, effectively removing the necessity of a coordination decision (since it is always a yes). Then after the support pairs are extracted, the cost between two neighbors is simply a constant.

\begin{algorithm}[t]
\caption{\textsc{Dynamic-hjsg} Search($\mathbb{G}_s, \nodeset_0, \nodeset_g$)}
\label{dynamichjsg}
\KwIn{Simplified graph $\mathbb{G}_s$, start joint-state $\nodeset_0$, goal joint-state $\nodeset_g$}
\KwOut{Shortest path $\pi$, per-agent goal costs $C_n$, total cost $C$ }
    \nl Initialize priority queue $pq \leftarrow [(0, \nodeset_0)]$ \\
    \nl $dist[\nodeset_0] \leftarrow 0$, $prev \leftarrow \emptyset$, $visited \leftarrow \emptyset$,
    $goalCosts \leftarrow \emptyset$ \\
    \nl \While{$pq$ not empty}{
        \nl $(d, {\bf{JS}}_i) \leftarrow$ pop from $pq$ \\
        \nl \If{${\bf{JS}}_i \in visited$}{
        \nl {\textbf{continue}}\\
        } 
        \nl Add ${\bf{JS}}_i$ to $visited$ \\
         \nl    \If{${\bf{JS}}_i$ is empty}{
         \nl \Return(path to ${\bf{JS}}_i$, $goalCosts$, $sum(goalCosts)$)\tcp*{Paths found}}
         \nl    \If{${\bf{JS}}_i \notin dist$ or $d < dist[{\bf{JS}}_i]$}{
                \nl $dist[{\bf{JS}}_i] \leftarrow d$, push $(d, {\bf{JS}}_i)$ to $pq$ \\
            }
        \nl \For{neighbor ${\bf{JS}}_j$ of ${\bf{JS}}_i$ from 2-agent moves}{
            \nl \If{${\bf{JS}}_j \in visited$}{\nl \bf{continue}} 
            \nl $c \leftarrow$ ComputeSuperEdgeCost($\mathbb{G}_s$, ${\bf{JS}}_i$, ${\bf{JS}}_j$) \tcp*{A constant after extracting support pairs}
            \nl \If{$c = \infty$}{\nl \bf{continue}} 
            \nl \If{${\bf{JS}}_j \notin dist$ or $dist[{\bf{JS}}_i] + c < dist[{\bf{JS}}_j]$}{
                \nl $dist[{\bf{JS}}_j] \leftarrow dist[{\bf{JS}}_i] + c$ \\
                \nl $prev[{\bf{JS}}_j] \leftarrow {\bf{JS}}_i$ \\
                \nl push $(dist[{\mathbf{JS}}_j], {\mathbf{JS}}_j)$ to $pq$\\
            }
        }
    }
    \nl\Return{$(\emptyset, goalCosts, \infty)$} \tcp*{Not all paths found}
\end{algorithm}
\textbf{\textsc{Dynamic-hjsg} Search.}
Alg.~\ref{dynamichjsg} describes the entire \emph{\textsc{Dynamic-hjsg} Search} procedure, on the simplified graph $\mathbb{G}_s$. 
The algorithm begins by initializing a priority queue with the start state (line 1), and sets up tracking structures for distances, predecessors, and visited states (line 2). It uses a Dijkstra-like search over the joint-state space, where each joint-state ${\bf{JS}}_i$ represents the current positions of all remaining agents.

Within the while loop (lines 3-22), for each step, it tries to explore the neighborhood of the frontier.
Their individual cost-to-goal is recorded as $goalCosts$. If all agents have reached their goals, the algorithm reconstructs the full solution path and returns it along with $goalCosts$ and the total cost (lines 8-9).
For joint-states where not all agents are completed, the algorithm explores neighboring states generated by 2-agent joint moves (lines 12-21). The edge cost between joint-states is computed (line 15) via the ComputeSuperEdgeCost function, which outputs a constant after extracting support pairs, as a result of lemma.~\ref{lemma1}. If a cheaper path to a neighbor is found, it updates the distance and predecessor mappings and pushes the neighbor into the priority queue (lines 18-21).
If the goal joint-state is unreachable, the algorithm returns an empty path, infinite cost, and any partial $goalCosts$ collected (line 22).
This dynamic pruning approach significantly reduces the combinatorial explosion of the full joint-state space, especially in environments with heterogeneous goal structures or when agents complete their tasks at different times.

Complexity:
Assume the max degree of each node in the simplified graph is $d$, the number of visited joint-states is $P$, and the number of visited joint-edges is $Q$. Then, for each joint-state, it has $O(N^2\cdot(d+1)^2) = O(N^2\cdot d^2)$ neighboring joint-states, since only two robots are allowed to move at a time, which implies $Q=O(P\cdot N^2\cdot d^2)$. The time complexity for computing the cost of each joint-edge (via Hungarian algorithm) is $O(N^3)$, so total edge cost computation is $O(Q\cdot N^3)$. Dijkstra’s algorithm runs in $O(Q\cdot\log P)$, since it expands $P$ states with $Q$ edges. Therefore, the total complexity is $O(Q\cdot N^3 + Q\cdot\log P) = O((P\cdot N^2\cdot d^2)\cdot (N^3 + \log P))$, where $P<<|\nodeset_s|^N$. This is exponentially better than the full JSG approach. In practice, $P$ is polynomial in $N$ for many sparse environments or goal-driven navigation, and thus practical in many cases.

\section{Experiment Results}
\label{sec::results}
We conduct experiments on a variety of graphs to evaluate the optimality and efficiency of \textsc{Dynamic-hjsg} against other proposed algorithms. 
Our setting varies as follow:\\
1) Variable Graph size: \# nodes = 6, 9, 12, \& 15. \\
2) Variable Graph types: $random$ where the edges are randomly generated, $rect\_perfect$ where the graph is a perfect grid, \& $voronoi$ that forms a voronoi diagram.\\
3) Fixed risk edge ratio:  0.2 (we make it fixed for now), which means 20\% of all edges are risky edges whose costs are higher than others, and possibly reduced to costs lower than others if support is offered.\\
4) Fixed support node number: 1 (we make it fixed for now), which means each risky edge has 1 support node. We use this setting because \textsc{ces} scales terribly with the number of support pairs.\\
5) Variable Agent number: \# agents = 2, 3, 4, 5, 6. We use this setting because the original \textsc{jsg} doesn't scale well with \# agents.\\
The average edge density is 0.3, producing moderate to dense graphs.
Each experiment is repeated on 3 different randomly generated graphs (random.seed(12)) for each setting on Mac M1. 
Additionally, we terminate experiments when the runtime is over 60s.

Becuase traditional 3DM solvers don't work with transition-dependent costs, we compare the proposed \textsc{Dynamic-hjsg} with baselines:\\
1) Original \textsc{jsg}: \textsc{Dynamic-hjsg} with no simplified graph, no dynamic joint-state construction, no limited neighborhood, and no incremental goal satisfaction.\\
2) Original \textsc{ces}: exhaustive search on every possible matching, with assumption that each support pair is used once.\\
3) Homogeneous \textsc{ces} (\textsc{hces}): modified \textsc{ces} with Floyd Warshall algorithm that calculates piecewise path cost in the beginning, so it doesn't have to calculate it in each iteration of the loops.\\
(\textsc{rhoca*} gives terrible results when the agents are spread out in the graph, and it is not evaluated. Plus: \textsc{Dynamic-hjsg} can be seen as \textsc{rhoca*} with horizon size adaptive to the graph structure)

\subsection{Results}
Both \textsc{Dynamic-hjsg} and \textsc{jsg} are optimal methods, and \textsc{hces} and \textsc{ces} are optimal if the assumption is not violated. Therefore, we mainly evaluate algorithm efficiency here.
\begin{figure}
    \centering
    \includegraphics[width=0.9\linewidth]{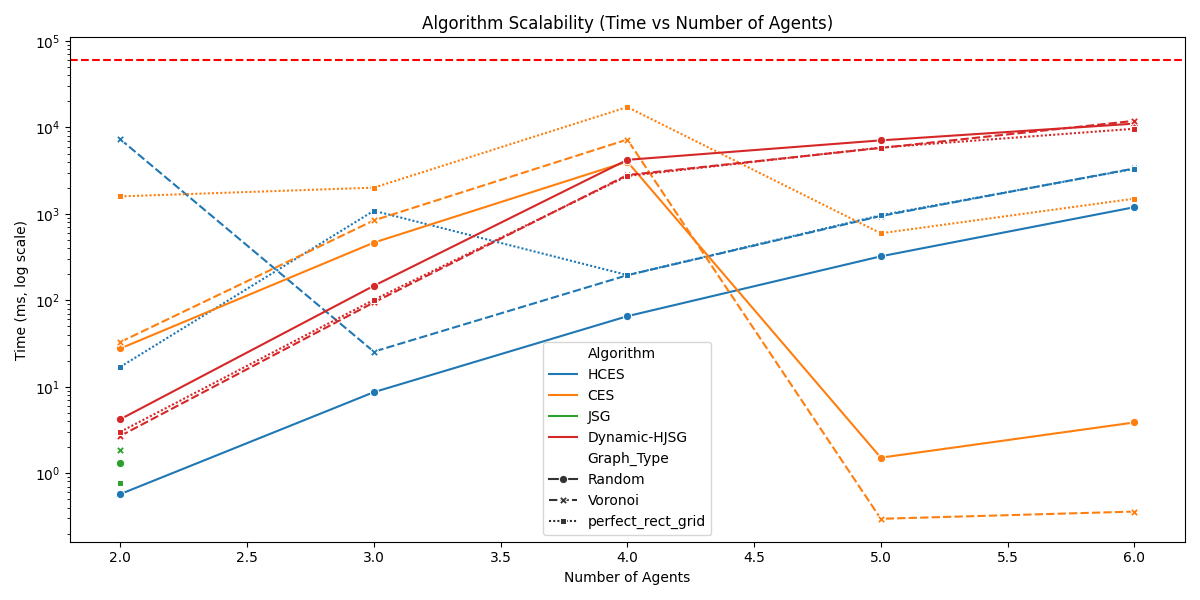}
    \caption{Runtime vs. \# Agents}
    \label{scalability}
\end{figure}


\begin{table}[H]
    \centering
    \small 
    \begin{tabular}{|c|c|c|c|c|}
    \hline
    Algorithm &Completed Runs (\%)	&Avg Runtime (Success)	&Timeout Cases	&Effective Runtime*\\
    \hline
    Dynamic-HJSG	&98\%	&120ms	&2\%	&132ms\\
    \hline
    HCES	&85\%	&450ms	&15\%	&9.45s\\
    \hline
    JSG	&65\%	&1.2s	&35\%	&42.2s\\
    \hline
    CES	&40\%	&3.8s	&60\%	&63.8s\\
    \hline
    \end{tabular}
    \caption{Runtime Analysis}
    \label{RuntimeAnalysis}
\end{table}


\subsubsection{Runtime Analysis}
\textsc{Dynamic-hjsg} demonstrated superior scalability, maintaining sub-second median runtime ($10^0-10^2$ ms) across all team sizes (Fig.~\ref{scalability}). The algorithm showed polynomial time complexity with respect to team size. \textsc{hces} exhibited moderate scalability, with runtime remaining below 10s for teams of up to 5 agents. Performance degraded sharply beyond this threshold (60s timeout reached in 15\% of 6-agent cases). Legacy algorithms from the conference version (\textsc{ces} and \textsc{jsg}) showed exponential time complexity, with:
\textsc{ces} exceeding 60s timeout in 60\% of 4+ agent cases
\textsc{jsg} timing out in 35\% of 5+ agent cases, as shown in Table.~\ref{RuntimeAnalysis}. 
Note that the curves in Fig.~\ref{scalability} do not reflect the real average runtime, due to the timeout of 60s (meaning high runtime is not recorded for some runs). But they do show the polynomial tendency wrt \# agents of the runtime of \textsc{Dynamic-hjsg} as it completes 98\% runs within the timeout. \textsc{ces} was already proven polynomial to \# agents in the conference version, but it has worse than exponential runtime growth wrt. \# support pairs, which is why its timeout rate is so high with the setting of 20\% risky edge rate. To our surprise, the modified version \textsc{hces} is performing so well, but that might be owing to the support node number of each risky edge is fixed at 1, which significantly limits \# support pairs. 

\begin{figure}
    \centering
    \includegraphics[width=0.9\linewidth]{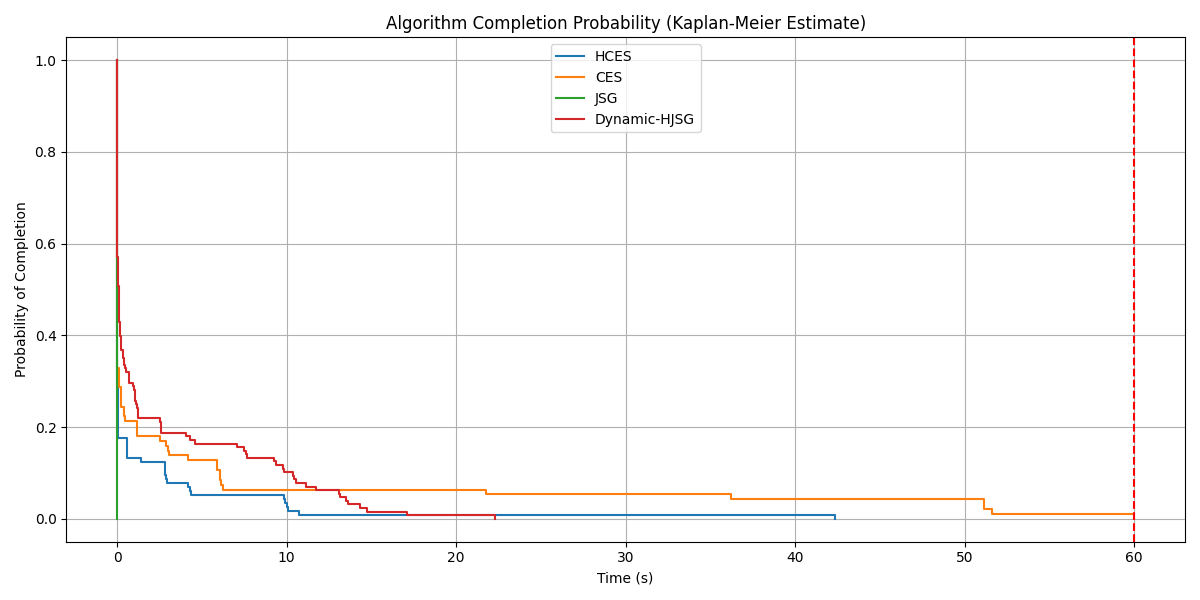}
    \caption{Algorithm Completion Probability}
    \label{AlgorithmCompletionRate}
\end{figure}

We also conduct a survival analysis as shown in Fig.~\ref{AlgorithmCompletionRate}, summarized in Table.~\ref{AlgorithmCompletionRate}. The analysis treats terminated executions as right-censored observations, preserving their information up to the point of timeout. Dynamic-HJSG demonstrates superior reliability with 98\% completion probability (95\% Confidence Interval (CI): 96-100\%) at 60 seconds, maintaining a near-linear survival curve that reflects consistent polynomial-time performance. In contrast, \textsc{ces} exhibits rapid decay with only 40\% completion probability (95\% CI: 35-45\%) by 60 seconds, characteristic of exponential time complexity. The hazard rates (instantaneous timeout risk) reveal critical inflection points: \textsc{hces} shows stable performance until agent counts exceed 5 (30-second hazard rate increase from 0.02 to 0.15), while \textsc{jsg}'s hazard rate grows monotonically, reaching 0.33 by 60 seconds on perfect\_rect\_grid problems. These results empirically validate that \textsc{Dynamic-hjsg}'s state-space pruning effectively controls timeout risk, whereas legacy methods become unreliable beyond 3 agents. The crossing survival curves between \textsc{jsg} and \textsc{hces} at 42.5 seconds (Log-rank test $p<0.001$) further confirm their fundamentally different scalability properties.

\begin{table}[h]
    \centering
    \small
    \begin{tabular}{|c|c|c|c|}
    \hline
         Algorithm	&30s Completion \%	&60s Completion \%	&Median Runtime\\
         \hline
        Dynamic-HJSG	&92\%	&98\%	&8.2s\\
        \hline
        HCES	&75\%	&85\%	&22.4s\\
        \hline
        JSG	&40\%	&65\%	&48.1s\\
        \hline
        CES	&25\%	&40\%	&Timeout\\
        \hline
    \end{tabular}
    \caption{Comparative Analysis of Completion Probability}
    \label{ComparativeAnalysis}
\end{table}

\section{Conclusions \& Discussion}
\label{conclusions}
In this paper, we present a systematic problem formulation and mathematical analysis of \textsc{tcgre}, proving its NP-hardness and demonstrating that efficient decomposition is key to solving this problem: dividing a 3D Matching problem into two 2D Matching sub-problems. We recapitulate the 3 classes of algorithms from the conference version, extract their essence, and present the \textsc{Dynamic-hjsg} algorithm for efficient multi-agent coordination on a compressed joint-state graph. The algorithm dynamically updates the joint goal states to a limited joint-state neighborhood, as individual agents reach their targets, enabling early termination and cost extraction. Our method effectively captures coordination benefits and accurately reflects the cooperative structure of the environment. Overall, \textsc{Dynamic-hjsg} offers a scalable and modular solution to coordination in homogeneous multi-agent systems under risk-aware constraints.

While originally designed for abstract multi-agent coordination, \textsc{Dynamic-hjsg} is naturally applicable to the Multi-Agent Path Finding (\textsc{mapf}) problem. In \textsc{mapf}, multiple agents must move from start to goal positions on a shared graph without collisions. Traditional approaches often struggle with the combinatorial explosion of the joint configuration space or fail to model implicit cooperation opportunities. \textsc{Dynamic-hjsg} addresses these challenges in several ways:\\
1) {\bf{Joint-State Compression}}: By grouping agent states and transitions into super joint-states and risky/support structures, \textsc{Dynamic-hjsg} reduces the effective search space compared to explicit joint-state expansions in mapf solvers.\\
2) {\bf{Flexible Coordination Modeling}}: The use of risky edges and support nodes generalizes conflict detection in \textsc{mapf}. Instead of binary collision checks, \textsc{Dynamic-hjsg} assigns costs to potential conflicts and reduces them through cooperative matching, which aligns well with the soft constraints seen in \textsc{mapf} variants such as \textsc{mapf} with congestion, time windows, or communication limits.\\
3) {\bf{Incremental Goal Satisfaction}}: \textsc{Dynamic-hjsg} supports agents reaching their goals asynchronously, which is beneficial in large-scale \textsc{mapf} scenarios where agents may finish at different times or transfer responsibility dynamically.\\
4) {\bf{Integration Potential}}: The algorithm can be integrated as a post-processing refinement for \textsc{mapf} plans (to exploit additional cooperation), or as a complete solver when planning over abstracted or clustered agent groups.



\section{Acknowledgments}


\bibliographystyle{ACM-Reference-Format}
\bibliography{JAIR_Example_Template}

\appendix

\section{Reproducibility Checklist for JAIR}

Select the answers that apply to your research -- one per item. 

\subsection*{All articles:}

\begin{enumerate}
    \item All claims investigated in this work are clearly stated. 
    [yes]
    \item Clear explanations are given how the work reported substantiates the claims. 
    [yes]
    \item Limitations or technical assumptions are stated clearly and explicitly. 
    [yes]
    \item Conceptual outlines and/or pseudo-code descriptions of the AI methods introduced in this work are provided, and important implementation details are discussed. 
    [yes]
    \item 
    Motivation is provided for all design choices, including algorithms, implementation choices, parameters, data sets and experimental protocols beyond metrics.
    [yes]
\end{enumerate}

\subsection*{Articles containing theoretical contributions:}
Does this paper make theoretical contributions? 
[yes] 

If yes, please complete the list below.

\begin{enumerate}
    \item All assumptions and restrictions are stated clearly and formally. 
    [yes]
    \item All novel claims are stated formally (e.g., in theorem statements). 
    [yes]
    \item Proofs of all non-trivial claims are provided in sufficient detail to permit verification by readers with a reasonable degree of expertise (e.g., that expected from a PhD candidate in the same area of AI). [yes]
    \item
    Complex formalism, such as definitions or proofs, is motivated and explained clearly.
    [yes]
    \item 
    The use of mathematical notation and formalism serves the purpose of enhancing clarity and precision; gratuitous use of mathematical formalism (i.e., use that does not enhance clarity or precision) is avoided.
    [yes]
    \item 
    Appropriate citations are given for all non-trivial theoretical tools and techniques. 
    [yes]
\end{enumerate}

\subsection*{Articles reporting on computational experiments:}
Does this paper include computational experiments? [yes]

If yes, please complete the list below.
\begin{enumerate}
    \item 
    All source code required for conducting experiments is included in an online appendix 
    or will be made publicly available upon publication of the paper.
    The online appendix follows best practices for source code readability and documentation as well as for long-term accessibility.
    [partially]
    \item The source code comes with a license that
    allows free usage for reproducibility purposes.
    [no]
    \item The source code comes with a license that
    allows free usage for research purposes in general.
    [no]
    \item 
    Raw, unaggregated data from all experiments is included in an online appendix 
    or will be made publicly available upon publication of the paper.
    The online appendix follows best practices for long-term accessibility.
    [no]
    \item The unaggregated data comes with a license that
    allows free usage for reproducibility purposes.
    [no]
    \item The unaggregated data comes with a license that
    allows free usage for research purposes in general.
    [no]
    \item If an algorithm depends on randomness, then the method used for generating random numbers and for setting seeds is described in a way sufficient to allow replication of results. 
    [partially]
    \item The execution environment for experiments, the computing infrastructure (hardware and software) used for running them, is described, including GPU/CPU makes and models; amount of memory (cache and RAM); make and version of operating system; names and versions of relevant software libraries and frameworks. 
    [yes]
    \item 
    The evaluation metrics used in experiments are clearly explained and their choice is explicitly motivated. 
    [yes]
    \item 
    The number of algorithm runs used to compute each result is reported. 
    [yes]
    \item 
    Reported results have not been ``cherry-picked'' by silently ignoring unsuccessful or unsatisfactory experiments. 
    [yes]
    \item 
    Analysis of results goes beyond single-dimensional summaries of performance (e.g., average, median) to include measures of variation, confidence, or other distributional information. 
    [yes]
    \item 
    All (hyper-) parameter settings for 
    the algorithms/methods used in experiments have been reported, along with the rationale or method for determining them. 
    [partially]
    \item 
    The number and range of (hyper-) parameter settings explored prior to conducting final experiments have been indicated, along with the effort spent on (hyper-) parameter optimisation. 
    [partially]
    \item 
    Appropriately chosen statistical hypothesis tests are used to establish statistical significance
    in the presence of noise effects.
    [yes]
\end{enumerate}

\subsection*{Articles using data sets:}
Does this work rely on one or more data sets (possibly obtained from a benchmark generator or similar software artifact)? 
[yes/no]

If yes, please complete the list below.
\begin{enumerate}
    \item 
    All newly introduced data sets 
    are included in an online appendix 
    or will be made publicly available upon publication of the paper.
    The online appendix follows best practices for long-term accessibility with a license
    that allows free usage for research purposes.
    [NA]
    \item The newly introduced data set comes with a license that
    allows free usage for reproducibility purposes.
    [no]
    \item The newly introduced data set comes with a license that
    allows free usage for research purposes in general.
    [no]
    \item All data sets drawn from the literature or other public sources (potentially including authors' own previously published work) are accompanied by appropriate citations.
    [NA]
    \item All data sets drawn from the existing literature (potentially including authors’ own previously published work) are publicly available. [NA]
    \item All new data sets and data sets that are not publicly available are described in detail, including relevant statistics, the data collection process and annotation process if relevant.
    [NA]
    \item 
    All methods used for preprocessing, augmenting, batching or splitting data sets (e.g., in the context of hold-out or cross-validation)
    are described in detail. [NA]
\end{enumerate}

\subsection*{Explanations on any of the answers above (optional):}
This paper focuses on solving a problem using traditional mathematical approach and is not data dependent. A thorough theoretical analysis and provided before we introduce the method, which is clearly described (if not too verbose), with detailed pseudocode, and baseline details were also provided in the conference version. It should be fairly easy to reconstruct everything. If necessary, I would upload a snippet on github upon publication.

\end{document}